\def\sqr#1#2{{\vcenter{\vbox{\hrule height.#2pt \hbox{\vrule
width.#2pt height#1pt \kern#1pt \vrule width.#2pt} \hrule height.#2pt}}}}
\def\square{\mathchoice\sqr54\sqr54\sqr{2.1}3\sqr{1.5}3}
\documentstyle[12pt]{article}
\begin{document}
\begin{titlepage}
\title{Structure of Primordial Gravitational Waves Spectrum in a Double
Inflationary Model}
\author{David Polarski$^{1,2}$ and Alexei A. Starobinsky$^{a,b}$\\
\hfill \\
$^1$~{\it Laboratoire de Mod\`eles de Physique Math\'ematique, EP93 CNRS}\\
{\it Universit\'e de Tours, Parc de Grandmont, F-37200 Tours (France)}\\
\hfill\\
$^2$~{\it D\'epartement d'Astrophysique Relativiste et de Cosmologie},\\
{\it Observatoire de Paris-Meudon, 92195 Meudon cedex (France)}\\
\hfill \\
$^a$~{\it Landau Institute for Theoretical Physics}, \\
{\it Kosygina St. 2, Moscow 117334 (Russia)} \\
\hfill \\
$^b$~{\it Yukawa Inst. for Theor. Physics, Kyoto University, Uji 611 (Japan)}}

\date{\today}
\maketitle

\begin{abstract}
The gravitational waves (GW) background generated in a double inflationary
model, with two scalar fields mutually interacting through gravity
only, and its relative contribution $T/S$ to large-angle temperature
fluctuations of the relic microwave background are investigated in detail.
The relation between $T/S$ and the slope of the GW spectrum
$n_T$ is shown to be a discriminative test between a slow-roll
inflation driven by one scalar field and more complicated models.
It is found that the GW amplitude
is not exactly zero in minima of spectral oscillations, this
property being an observational, in principle, manifestation
of GW being in a squeezed vacuum state during inflation.
\end{abstract}

\end{titlepage}

Cosmological models in which the early Universe passed through two
successive de Sitter (inflationary) phases of quasi-exponential
expansion (\cite{lev85}-\cite{polarski} and other papers) are usually
called double
inflationary ones. The initial spectrum of adiabatic perturbations in
these models is not approximately flat (i.e., $n\approx 1$, or
Harrison-Zeldovich like), it has a characteristic scale and more power
on larger scales. This represents one of the possible ways to reconcile
the cold dark matter model (CDM) with observations of large-scale spatial
distributions of galaxies and their peculiar velocities~\cite{mad90}
and large-angle fluctuations of the relic microwave background
temperature~\cite{smoot92}, because the observational data imply an
excess of power in the
perturbation spectrum on scales beginning from $(25-50)h^{-1}$Mpc
up to the present Hubble radius $R_H=cH_0^{-1}=3000h^{-1}$Mpc as compared
to the CDM model with flat initial spectrum of adiabatic
perturbations (the standard CDM model) normalized to produce a correct
galaxy-galaxy correlation function and having the biasing parameter
$b\sim 2$ on a scale of $8h^{-1}$Mpc.

The most logically economic double inflationary models possess only
2 additional free dimensionless parameters as compared to the standard
CDM model. Therefore, they still retain most of the predictive
power of the inflationary scenario. Detailed comparison of a number of
double inflationary models with recent observational
data~\cite{stefan93,pps} shows
that they are compatible with observations (though for a rather small
window of parameters). In~\cite{polarski}, we performed a detailed
investigation
of a simple double inflationary model with two massive scalar fields
and found initial spectra of adiabatic perturbations and
gravitational waves generated from vacuum quantum fluctuations
during a double inflationary stage analytically in the case when
there is a pronounced intermediate stage between two stages of inflation.
As explained in~\cite{pps}, power spectra for which the relative height of the
two "plateaus" in the spectrum of adiabatic perturbations is
compatible with observations correspond to values of
${{m_h}\over {m_l}}\approx 10-15$, the latter number increases to 18 if the
recently proposed fit to the COBE data corresponding to $Q_{rms-PS}=20~\mu K$
for $n=1$~\cite{gorski} is used (this is only a little bit less than 15-25
that was proposed in~\cite{polarski} before the announcement of the COBE
measurements). The intermediate stage between the two inflationary
stages is not pronounced in that case. Therefore we have to resort to
numerical simulations in order to get the perturbation spectra with good
accuracy.
A unique and remarkable property of some variants of the inflationary scenario
(in particular the chaotic and power-law inflations) is that such a large
primordial gravitational wave background is generated from vacuum quantum
fluctuations that it is fairly reasonable that a modest, but still significant
part of the large-angle fluctuations of the cosmic microwave background (CMB)
temperature $\frac{\Delta T(\theta,~\varphi)}{T}$ observed by
COBE~\cite{smoot92} is due to these
gravitational waves. That is why it is very important to calculate the ratio
of the gravitational waves amplitude to that of scalar perturbations and their
relative contribution to the temperature anisotropy in the double inflationary
scenario.

In this letter, we extend this analytical and numerical study to the
spectrum of primordial gravitational waves and their contribution to
the large-angle
temperature fluctuations of the CMB. Though
the ratio of GW amplitude to that of adiabatic
perturbations and the corresponding ratio of their contributions
to the large-angle $\Delta T/T$ fluctuations appear to be rather small
in the case under consideration, this study gives a possibility to
obtain important basic conclusions about a
discriminative observational test between slow-roll inflation
driven by one scalar field and more complicated models, in particular double
inflation, and about an observable (in principle) feature
indicating that GW were in a pure squeezed state for some time during
inflation.

Let us give the essentials of our model of double inflation and generation
of perturbations in it.
The phenomenological Lagrangian density describing matter and
gravity during and near the inflationary stage, is given by
\begin{equation}
L=-{R\over {16\pi G}}+{1\over 2}(\phi_{h,\mu}\phi_h^{,\mu}-m_h^2
\phi_h^2) +{1\over 2}(\phi_{l,\mu}\phi_l^{,\mu}-m_l^2 \phi_l^2)
\end{equation}
where $\mu=0,..,3, c=\hbar=1, m_h\gg m_l$ and the Landau-Lifshitz sign
conventions are used. The space-time metric has the form
\begin{equation}
ds^2=dt^2-a^2(t)\delta_{ij}dx^i dx^j, \qquad\ i,j=1,2,3.
\end{equation}
Spatial curvature may always be neglected because it becomes vanishingly small
during the first period of inflation driven by the heavy scalar field.
The background fields are homogeneous and classical.
The perturbed FRW background, in longitudinal gauge, has a metric
given by
\begin{equation}
ds^2=(1+2\Phi)dt^2-a^2(t)(1-2\Psi)\delta_{ij}dx^idx^j~.
\end{equation}
The same mechanism that produces quantum fluctuations of the scalar fields
will also produce tensor metric perturbations or gravitational waves. For each
of the polarization states $\lambda$, where $\lambda=1,2$, and the
polarization tensor is normalized by the condition $e_{ij}e^{ij}=2$, the
amplitude $h_{\lambda}$ is given by
\begin{equation}
h_{\lambda}=\sqrt{16\pi G}\phi_{\lambda}
\end{equation}
where $\phi_{\lambda}$ is a massless scalar field satisfying
\begin{equation}
\square~ \phi_{\lambda}=0\label{GW}.
\end{equation}
Let us consider the spectrum of tensor and scalar adiabatic perturbations
\begin{equation}
h(k)=Bk^{-3/2}~,\qquad\ \Phi(k)={3\over 10}Ak^{-3/2}~.
\end{equation}
The quantities $A$ and $B$ are those introduced in~\cite{star83,starGW}, they
are only weakly depending on $k$ while $h(k)$ and $\Phi(k)$ are defined as
follows
\begin{equation}
\langle h_{{\bf k},\lambda }, h^*_{{\bf k'},\lambda '}\rangle \equiv {1\over
 2}h^2(k)
\delta^{(3)}({\bf k}-{\bf k'})\delta_{\lambda \lambda '}~,
\end{equation}
\begin{equation}
\langle \Phi_{\bf k}\Phi^*_{\bf k'}\rangle =
\Phi^2(k)\delta^{(3)}({\bf k}-{\bf k'})
\end{equation}
where the Fourier transform  conventions are
$\Phi_k\equiv {1\over{(2\pi)^{3\over 2}}}\int \Phi({\bf
r})e^{-i{\bf kr}}d^{3}{\bf k}$. We assume also that $a(t)\propto t^{2/3}$
today.

Let us first consider the ratio of the amplitudes of GW and adiabatic
perturbations generated in double inflation.
For wavelengths well outside the Hubble radius, ${k\over a}\ll H$,
each Fourier mode $h_{k,\lambda}(\eta)$ has a constant value. If they
cross the Hubble radius during a quasi-de Sitter stage, $h(k)$ is given by
\begin{equation}
h(k)=\sqrt{16\pi G}{H(t_k)\over {\sqrt{k^3}}}\label{ds}
\end{equation}
where $t_k$ is the
Hubble radius crossing time defined as ${k\over {a(t_k)}}=H(t_k)$.
In (\ref{ds}) both polarization states are taken into account.
Note that long wavelengths
cross the Hubble radius towards the end of the first
inflation, so that $H(t_k)=\sqrt{{2\over 3}m_h^2 \ln {{k_b}\over k}}$ in
agreement with our numerical simulations
\begin{equation}
k^{3\over2}h(k)=\sqrt{{{32\pi G m_h^2}\over 3}}\sqrt{\ln {{k_b}\over k}}=
\sqrt{{{32\pi G m_h^2}\over 3}}\sqrt{s(k)-s_0},\hspace{1cm}s>s_0
\end{equation}
where $k_b$ is the characteristic scale of the spectrum corresponding
to the end of the first inflation driven by $\phi_h$,
$s=2\pi G(\phi_l^2+
\phi_h^2)$ is the number of e-folds till the end of the second inflation,
$s(k)\equiv s(t_k)=\ln \frac{k_f}{k} = \ln \frac{a_f}{a}$, $k_f=a_f H_f$
where the index $f$ denotes the values
of all quantities at the end of the second inflation, and
$s_0\gg 1$ is the total number of e-folds during the second inflation. We
take $s_0=2\pi G\phi_0^2
\approx 60$, in order to get the characteristic scale $k_b$ at
the right place today. For the scalar
adiabatic perturbations we have~\cite{polarski}
\begin{eqnarray}
k^{3\over2}\Phi(k) =  {{6\sqrt{2}\pi G}\over 5}\Bigl (
H\sqrt{\phi_l^2+\phi_h^2}\Bigr )_{t_k}
= {{\sqrt{24\pi G m_h^2}}\over 5}\sqrt{s(k)(s(k)-s_0)}~,\hspace{1cm}s>s_0~.
\end{eqnarray}

As mentioned above, we are mostly interested in those values of
$\frac{m_h}{m_l}$ for which the intermediate stage, though it exists, is not
well
pronounced and deviations from monotonous behaviour of $h(k)$ and $\Phi (k)$
are concentrated in the region $|s-s_0|\sim 1$.
If the Hubble radius is crossed
during the second inflation, we obtain the usual result for inflation
driven by one scalar field
\begin{equation}
k^{3\over2}h(k)= \sqrt{{{32\pi G m_l^2}\over 3}}\sqrt{s(k)}~, \qquad\
k^{3\over2}\Phi(k)={{\sqrt{24\pi G m_l^2}}\over 5}s(k)~,\hspace{1cm}s<s_0~.
\end{equation}
This finally yields
\begin{equation}
R(k)\equiv {h(k)\over \Phi (k)}={5\sqrt 2 \over 3\sqrt{\pi G}
(\sqrt{\phi_l^2+\phi_h^2})_{t_k}}=
{10 \over 3\sqrt{s(k)}} \label{R}
\end{equation}
for both $s>s_0$ and $s<s_0$, aside from the region $|s-s_0|\leq s_0
\frac{m_l^2}{m_h^2}\sim 1$ around the point $s=s_0$. For $s\approx s_0$,
$R(k)\approx 0.43$.
The behaviour of $R(k)$ obtained numerically
(see Figs.2-3) is in excellent agreement with (\ref{R}).
The slope of $R(k)$ is:
\begin{equation}
{{d\ln R(k)}\over {d\ln k}} \equiv \frac{1}{2}(n_T-n_S+1)
={1\over 2s(k)} \ll 1\label{slopeR}
\end{equation}
where $n_S\equiv 1+\frac{d\log(k^3\Phi^2(k))}{d\log k},~n_T
\equiv \frac{d\log(k^3
h^2(k))} {d\log k}$.  In accordance with~(\ref{slopeR}), we see from Fig.1
that $h(k)$ has a shape very similar to that of
$\Phi(k)$. However, this is not surprising since we have shown
numerically~\cite{david} that the contribution
of the light scalar field perturbations to $\Phi(k)$ is dominant for the
values of ${{m_h}\over {m_l}}$ considered here, while on the
other hand the equation governing the light scalar field fluctuations
essentially coincides with eq.~(\ref{GW}), the equation for gravitational
waves.

We see that the relation
\begin{equation}
n_T\approx n_S-1\label{n}   \label{NS}
\end{equation}
(proposed, e.g., in~\cite{critt}) which has a very narrow region of
applicability even for the case of
single inflation, namely it is valid for power-law inflation only while it is
not valid for chaotic inflation (see also the discusion in~\cite{kolb}),
turns out to be valid for double inflation for large scales satisfying
$k\ll k_b\equiv k_f e^{-s_0}$ (and not too small),
as can be seen from~(\ref{slopeR}) and also from the figures.
As shown in~\cite{pps}, the best fit to observational
data is given by $2\pi k_b^{-1}\approx (6-10)h^{-1}$Mpc, so the condition
$k\ll k_b~(s(k)>s_0)$ is certainly satisfied for scales probed by
CMB anisotropy experiments on large and intermediate angles.
On smaller scales for which $k>k_b$,
the relation~(\ref{n}) is not valid any more. Hence, it cannot
be considered as a test neither for generic inflation nor even for a single
inflation. More general is the following relation
which is valid for any slow-roll inflation driven by one scalar field:
\begin{eqnarray}
\frac{T}{S} \equiv \frac{\langle |a_{lm}|^2\rangle_{GW}}
{\langle |a_{lm}|^2\rangle_{AP}}={\cal K}_l\left(\frac{3}{10} R(k)
\right)^2={\cal K}_l |n_T|~,~~~n_T<0~,  \label{TS} \\
{\cal K}_l={\rm const}=\frac{25}{9}\left(1+\frac{48\pi^2}{385}\right)
\simeq 6.20~,\qquad\ l\gg 1~,~~~|n_T|\ll 1~, \label{TS1} \\
{\cal K}_2\approx 6.93,~~~{\cal K}_3\approx 5.45,~~~{\cal K}_5
\approx 5.10,~~~{\cal K}_{10}\approx 5.30,~~~|n_T|\ll 1  \label{TS2}
\end{eqnarray}
with ${\cal K}_5$ being the minimal value for integer $l<40$.
In these expressions, only the Sachs-Wolfe effect is taken into account
for both types of perturbations.
Eq.~(\ref{TS1}) was first derived in~\cite{starGW}, and~(\ref{TS2}) follows
from results obtained in~\cite{fabbri,starGW} (see also recent calculations
in~\cite{allen}). However, the quadrupole was not detected by COBE
and it is usually excluded from the data; the asymptotic value (\ref{TS1})
is not reached either for multipoles $l=3-15$ measured by COBE.
Taking into account the Doppler and the Silk effects for adiabatic
perturbations
yields a further decrease of ${\cal K}_{10}$ by $\approx 10\%$.
Thus, it is better to use the value ${\cal K}_l \sim 5$ when estimating
the GW contribution to the actual COBE data.

The relation~(\ref{TS}) is strongly violated in the case of double
inflation for $s(k)>s_0$ because we then have:
\begin{equation}
\frac{T}{S}= {\cal K}_l(0.3R(k))^2={\cal K}_l (s(k))^{-1},~~~
|n_T|=(s(k)-s_0)^{-1},  \label{nT}
\end{equation}
therefore, $T/S \ll {\cal K}_l|n_T|$. Note that relation~(\ref{TS1}) is not
valid either for the $R+R^2$ inflationary model
{}~\cite{star83,starGW}. So,~(\ref{TS}) represents a decisive test for single
inflation driven by a scalar field. Still this test is difficult because
it first requires identification of the GW contribution to the
${\Delta T\over T}$
fluctuations. Also, it cannot be used to predict the amount of this
contribution using the total COBE data since both sides of
relation~(\ref{TS}) (in contrast to~(\ref{NS})) contain quantities
referring to GW.

 For $s\approx s_0$, we get $T/S={\cal K}_{10}s_0^{-1}\approx 0.08$.
As contributions to the ${\Delta T\over T}$ fluctuations from gravitational
waves and adiabatic perturbations are statistically independent, the total
{\it rms} value of ${\Delta T\over T}$ at large angles gets increased by a
factor $\sqrt{1+(T/S)}$, i.e. by about 4\%, due to the GW background.
Therefore, in our model the ${\Delta T\over T}$ anisotropy on scales probed
by COBE comes mainly from adiabatic perturbations. However, when comparing
predictions of the model with observations~\cite{pps}, the GW
contribution has to be taken into account.

Now we consider the dip in the $R(k)$ curve (see Figs.2-4) lying in the
transition region between the two "plateaus" that arises for ${m_h\over m_l}
\ge 15$, i.e. when the intermediate period between two phases of inflation
is sufficiently pronounced. Though, as can be seen from the figures,
it constitutes a small effect even for ${m_h\over m_l}=25$,
it is very interesting theoretically for two reasons. First, its very
existence is due to the effect discovered in~\cite{polarski}: the
necessity of taking into account a contribution of the heavy scalar field
quantum fluctuations
to the adiabatic perturbations in this range of scales.
Indeed, in the absence of break between two phases of inflation
(${m_h\over m_l}<15$), quantum fluctuations of the light scalar field
produce the dominant
contribution to adiabatic perturbations over the region involved since
$\phi_l=\phi_0\gg \phi_h \sim 0.5{\rm M_p}$.  But if there exists a pronounced
intermediate power-law stage ($s_0m_l^2m_h^{-2}\ll 1$), the
contribution of the heavy scalar field fluctuations becomes important in the
transition region between the upper "plateau" and the lower one. It is
given by the second term in the r.h.s. of eq.(4.18) of ref.~\cite{polarski}:
\begin{eqnarray}
k^{3\over 2}\Phi(k)={{3\sqrt{3\pi Gm_h^2}}\over
{5\sqrt{2}}}\ln^{1/2}{{k_b}\over k}\Bigl (2\pi G\phi_0^2
\bigl ({{k_1}\over k}\bigr )^6 \sin^2\bigl (\sigma{k\over {k_1}}\bigr)
+(2\pi\sigma_1)^2({{k_1}\over k}\bigr )^2\ln {{k_b}\over k}\Bigr)^{1\over2},
\nonumber \\
\sigma_1 \approx 0.2875,~~\sigma \approx 2.804,~~k_1\simeq k_b
\left({s_0m_l^2\over m_h^2}\right)^{1/6},~~k_1\stackrel{<}{\sim}k
\stackrel{<}{\sim}k_b~,   \label{paper}
\end{eqnarray}
see eq. (4.32) in~\cite{polarski} for $k>k_b$. On the other hand
(see~\cite{slava91,polarski} and Fig.18.7, p.312 in~\cite{slava92}),
\begin{equation}
k^{3/2}h(k)=\sqrt{6\pi Gm_h^2\ln({k_b\over k})}
\left(k_1\over k\right)^3|\sin (\sigma{k\over k_1})|,~~~
k_1 \stackrel{<}{\sim} k \stackrel{<}{\sim}k_b~. \label{GWosc}
\end{equation}

The first, oscillating, term in the r.h.s. of~(\ref{paper}) arises from
the light scalar field fluctuations and is proportional to the r.h.s.
of~(\ref{GWosc}), while the second does not.
In the absence of the heavy scalar field contribution, $R(k)$ would be a smooth
function of $k$.  But the second term becomes
important near the points where $\sin \left(\sigma{k\over k_1}\right)=0$, or
$k=1.12k_1$
preventing $\Phi(k)$ from going to zero in contrast to $h(k)$. This
explains why $R(k)$ has a dip. For the values of
$\frac{m_h}{m_l}$ considered here, this effect is not big and can be calculated
quantitatively with numerical methods only. Neverhteless, the appearance of
the dip in the $R(k)$ curve clearly shows that the heavy field contribution
is present even if definitely subdominant.

A second new and remarkable point is that the minimum of the dip
is not exactly zero as follows from the previously known eq.(\ref{GWosc}).
This is a direct consequence of the GW background being produced from the
initial vacuum state during the first de Sitter stage via the process of
particle creation, or in other words, of the GW being in a squeezed vacuum
quantum state during inflation. The mathematical expression of this fact is
that the amplitude $h_{\lambda}(k)$ of the GW (operator) field modes has
the following form in the regime $k\ll aH$:
\begin{eqnarray}
h_{\lambda}(k) = \sqrt{16\pi G}\left(
C_1+C_2\int_{\infty}^{t}\frac{dt}{a^3(t)} \right)~,~~~C_1 =
\frac{H(t_k)}{\sqrt{2k^3}}~,~~~C_2=-\frac{ik^{3/2}}{\sqrt{2}H(t_k)}~.
\label{GW11}
\end{eqnarray}
The first term in~(\ref{GW11}) is the quasi-isotropic mode which was made
purely real by a time-independent phase rotation, the second one is
the decaying mode. The condition $C_1\Im C_2=-\frac{1}{2}$ expressing
quantum coherence between the two modes reflects the fact that initially,
for $t\to -\infty,~k\gg aH$, the quantum state of the mode was the vacuum.
Both terms in the right-hand side of eq.~(\ref{GW11}) are of the same order
at the Hubble radius crossing during the first inflation. After that, the
decaying mode quickly becomes exponentially small. If the latter mode
is neglected as is
usually done, then quantum coherence is lost and the resulting state of
the GW for given $\bf k$ becomes equivalent to a classical wave with a Gaussian
stochastic amplitude, a fixed temporal phase and some special correlation
between $\bf k$ and $-\bf k$ modes (see the
detailed discussion in~\cite{david3}). Actually, the quantum coherence
is most probably lost shortly after the end of the second inflation due to
interactions of the decaying mode with the environment though, as explained
in~\cite{david3}, there is no necessity to consider these interactions
explicitly. As a result, the present GW background is not in a squeezed vacuum
quantum state, it is purely classical. However, it is natural to assume that
quantum coherence is not lost throughout inflation, at least if the decaying
mode is not negligible.
\par
Then, for the double inflationary model, the decaying mode appears to be
non-negligible over some range of $k$ close to $k_b$ due to the existence of
a break between the two inflationary stages which yields two more Hubble
radius crossings for modes belonging to that range. Keeping both terms
in~(\ref{GW11}), in contrast to previous calculations~\cite{slava91,polarski},
and using the method developed in~\cite{polarski}, we obtain the following
result for $h_{\lambda}(k)$ after the last Hubble radius crossing during the
second inflation (in the limit $ s_0m_l^2m_h^{-2}\ll 1$):
\begin{eqnarray}
h_{\lambda}(k) &=& D_1 \sin (\frac{k}{k_1}\sigma)+iD_2 \cos (\frac{k}{k_1}
\sigma)~,\nonumber \\
D_1 &=& -\frac{3}{4}\left(\frac{k_1}{k}\right)^3\frac{H(t_k)}{\sqrt{2k^3}},~~~
D_2=-\frac{2\sqrt{2}~H_0^2~k^{3/2}}{3~H(t_k)~k_1^3}\label{h}~.
\end{eqnarray}
where $H_0$ is the Hubble parameter at the beginning of the second inflation.
Note that $h_{\lambda}(k)$ could be made purely real had we chosen to make
a different time-independent phase rotation.
\par
It is clear that the minimum of the dimensionless quantity $k^{3/2}h(k)$,
though small, is not zero and it is reached at
$k\approx \pi\frac{k_1}{\sigma}$:
\begin{eqnarray}
k^{3/2}h_{min}(k) &=& \sqrt{32\pi G} k^{3/2}|D_2|=\frac{16\pi^3~\sqrt{\pi G}~
H_0^2}{3~\sigma^3~H(t_k)}\approx \frac{10~m_l^2~s_0}{m_h {\rm M_p}}~,
\nonumber\\
\frac{h_{min}(k)}{h_{max}(k)} &=& \frac{8\sqrt{2}}{9}\left(\frac{k}{k_1}
\right)^6\frac{H_0^2}{H^2(t_k)}=\frac{8\sqrt{2}\pi^6}{9\sigma^6}\frac{m_l^2s_0}
{m_h^2\ln \left(\frac{\sigma k_f}{\pi k_1}\right)}\approx 2.5\frac{s_0m_l^2}
{m_h^2}~.\label{min}
\end{eqnarray}
Hence, though the mode $h_{\lambda}(k)$ is semiclassical after inflation, we
may say that there appears a "quantum signature" in the present spectrum
of the GW background due to the fact that the decaying mode should be taken
into account in order to derive the correct form of the spectrum.
\par
Eq.(\ref{min}) is in very good agreement
with the numerical simulation displayed on Fig.1, where the dimensionless
quantity $k^{3/2}h(k)$ is expressed in $m_h/{\rm M_p}$ units. For $p=25$, the
value of the minimum $\approx 0.2$ which yields the value $\frac{k}{k_1}
\approx 1.14$, in remarkable agreement with~(\ref{min}). Keeping the decaying
mode in the scalar perturbations too, we get for $R_{min}$:
\begin{eqnarray}
R_{min} = \frac{10}{3\sqrt{s_0}}\frac{1}{\sqrt{1+\left(\frac{9\pi \sigma_1}
{8}\frac{H^2(t_k)\phi_h(t_k)}{H_0^2 \phi_0}(\frac{k_1}{k})^4\right)^2}}
= \frac{10}{3\sqrt{s_0}}\frac{1}{\sqrt{1+\left(\frac{9\sigma_1\sigma^4~m_h^2~
\ln^{3/2}(\frac{k_1}{k})}{8\pi^3~m_l^2~s_0^{3/2}}\right)^2}}\label{min1}
\end{eqnarray}
for the first minimum $k=k_1\frac{\pi}{\sigma}$. The value of $\frac{h_{min}}
{h_{max}}$ and $R_{min}$ in the higher minima $k=nk_1\frac{\pi}{\sigma},~n>1$,
quickly grows and the oscillations become unnoticeable.
\par
Thus, the spectrum $h^2(k)$ has no zeroes. Note that the
necessity of taking into account both modes resulting in the absence of zeroes
(despite the existence of deep wells) in the perturbation spectrum may
arise even for adiabatic perturbations generated during
inflation driven by one scalar field, provided the slow-rolling condition
breaks for a short period of time, e.g. near values of the scalar field
where the first derivative of its potential experiences a fast change
\cite{star92}. So, we can formulate the following {\it quantum
positivity} principle: if perturbations are generated by the process of
particle creation from the vacuum, their power spectrum $P(k)$ is non-zero
for all modes. Of course, the reverse statement that the absence of zeroes
implies that the fluctuations arise from quantum vacuum fluctuations, is not
true. But the appearance of zeroes in a power spectrum implies the classical
origin of perturbations, and the acoustic (Sakharov) oscillations in a purely
baryon-radiation Universe is an example of a classical process leading to a
power spectrum with zeroes for some values of $k$.

\vskip 25pt

\leftline{\bf Acknowledgments}
\par\noindent
This work was started when one of the authors (A.S.) was a visiting
professor in the Yukawa Institute for Theoretical Physics, Kyoto
University, and finished when he was visiting France under the agreement
between the Landau Institute for Theoretical Physics and Ecole Normale
Sup\'erieure, Paris. A.S. thanks YITP, ENS and EP93 CNRS (Tours) for
financial support, Profs. Y. Nagaoka and J. Yokoyama for their
hospitality in YITP and Profs. E. Brezin and C. Barrabes for the
hospitality in ENS and Universit\'e de Tours respectively. A.S. was
also supported in part by the Russian Foundation for Basic Research,
grant 93-02-3631.

\vfill\eject
\begin{figure}
\caption{The quantities $\log k^{3/2}\Phi(k)$ (upper curve) and
$\log k^{3/2}h(k)$ (bottom curve), with the dimensionless quantities
$k^{3/2}\Phi(k)~{\rm and}~k^{3/2}h(k)$ expressed in ${{m_h}\over {\rm M_p}}$
units, are displayed as functions of $\log k$ ($k$scale is arbitrary) for
${{m_h}\over {m_l}}=25$, $k\ll aH$. For the upper curve $a(t)\propto t^{2/3}$
nowadays is assumed. We note the presence of an oscillation
in the transition region. The minimum of $k^{3/2}h(k)$ is not zero due to the
presence of the decaying mode as explained in the text.
We note also the similarity of both curves for the $k$-range displayed here
corresponding to wavelengths that cross the Hubble radius towards the end of
the first inflation (upper "plateau") and at the beginning of the second
inflation (lower plateau).}
\end{figure}

\begin{figure}
\caption{The quantity $\log R(k)\equiv \log {{h(k)}\over {\Phi(k)}}$ is
displayed as a
function of $\log k$ for ${{m_h}\over {m_l}}=25$. The $k$ scale is the same as
in Fig.1. A dip is clearly seen in the transition region due to the
contribution of the heavy scalar field fluctuations (see text). Note again
that the bottom of the dip is not zero.}
\end{figure}

\begin{figure}
\caption{The quantity $\log R(k)\equiv \log {{h(k)}\over {\Phi(k)}}$ is
displayed as a function of $\log k$ for: a)~${{m_h}\over {m_l}}=20$~~~b)
${{m_h}\over {m_l}}=15$. Again, a dip is seen which
decreases with decreasing ${{m_h}\over {m_l}}$, due to the fact that for
decreasing values of ${{m_h}\over {m_l}}$, the intermediate stage becomes
less pronounced and the contibution of the heavy scalar field fluctuations
decreases in the transition region.}
\end{figure}

\end{document}